# Heteroepitaxial integration of ZnGeN$_2$ on GaN buffers using molecular beam epitaxy

*M. Brooks Tellekamp[1],\*, Celeste L. Melamed[2,1], Andrew G. Norman[1], Adele Tamboli[1,2]*

[1] National Renewable Energy Laboratory, Golden, Colorado 80401, USA

[2] Department of Physics, Colorado School of Mines, Golden, Colorado 80401, USA

ABSTRACT

Recently theorized hybrid II-IV-N$_2$ / III-N heterostructures, based on current commercialized (In,Ga)N devices, are predicted to significantly advance the design space of highly efficient optoelectronics in the visible spectrum, yet there are few epitaxial studies of II-IV-N$_2$ materials. In this work, we present heteroepitaxial ZnGeN$_2$ grown on GaN buffers and AlN templates. We demonstrate that a GaN nucleating surface is crucial for increasing the ZnGeN$_2$ crystallization rate to combat Zn desorption, extending the stoichiometric growth window from 215 °C on AlN to 500 °C on GaN buffers. Structural characterization reveals well crystallized films with threading dislocations extending from the GaN buffer. These films have a critical thickness for relaxation of 20 nm – 25 nm as determined by reflection high energy electron diffraction (RHEED) and cross-sectional scanning electron microscopy (SEM). The films exhibit a cation-disordered wurtzite structure, with lattice constants $a$ = 3.216 Å ± 0.004 Å and $c$ = 5.215 Å ± 0.005 Å determined by RHEED and X-ray diffraction (XRD). This work demonstrates a significant step towards the development of hybrid $ZnGeN_2$-GaN integrated devices.

\* brooks.tellekamp@nrel.gov

Introduction

Compound semiconductors (i.e. III-V and II-VI compounds) have enabled all modern optoelectronics, including diode lasers, light emitting diodes (LEDs), sensors, and high-efficiency photovoltaics. These compounds allow the device engineer to select various materials based on lattice constant, band gap, band alignment, and electrical properties enabling precise control of the injection, confinement, and extraction of carriers. An additional degree of freedom is obtained through continuously miscible alloying of binary compounds, such as the GaAs-AlAs alloy system. However, the crystal quality requirements for optically active materials necessitate the minimization of defects, limiting useful materials to those with a lattice constant within approximately ± 1 – 2 % of an available substrate. Thus, despite their success, III-V materials such as the (Al,Ga,In)(As,P,Sb) system and the (Al,Ga,In)N system are significantly limited by dislocations originating from lattice mismatch between heterolayers. In the absence of a suitably lattice-matched substrate or active material, complex buffer schemes and metamorphic structures are used to limit defects to an acceptable level (approximately < $10^6$ $cm^{-2}$ for traditional III-V, < $10^8$ $cm^{-2}$ for III-N), which often add significant complexity and cost.

The traditional lattice constant–band gap constraint can be overcome by moving to a ternary system where, for example, two group III elements are replaced by one group II and one group IV element, maintaining the overall charge balance as well as a crystal structure similar to the base III-V structure (wurtzite for III-N, zincblende for III-As,P,Sb). This II-IV-$V_2$ system enables two additional degrees of freedom when tuning the band gap energy: alloying on the second cation site, and long-range ordering of the cation lattice sites. Cation ordering is well studied in III-V alloys like (Ga,In)P, however the band gap tuning effect is small (~0.2 eV) due to isovalent substitutions.[1–5] In contrast, heterovalent substitutions in II-IV-$N_2$ materials are predicted to induce ~ 1 eV of band gap tunability along a single lattice-constant line.[6]

Currently studied II-IV-$V_2$ materials consist of (Zn, Mg, Cd) on the group II site, (Si, Ge, Sn) on the group IV site, and (N, As, P) on the group V site. II-IV-nitrides form a cation-ordered orthorhombic supercell of the base wurtzite III-N structure.[6] When cation-disordered, II-IV-$V_2$ materials are structurally similar to the base wurtzite structure with cation site occupancies of 50%. The InN analog $ZnSnN_2$, with a cation-ordered band gap of 1.5 eV and a cation-disordered band gap of ~1 eV,[7] has received significant attention recently as a low-cost photovoltaic absorber.[8,9] $ZnGeN_2$, the focus of this work, is a GaN analog with a theoretical band gap of 3.6 eV[6,10] (also reported theoretically as 3.42 eV[11], and experimentally at room temperature as 3 eV - 3.4 eV[12–16]) and wurtzite lattice constant 3.22 Å[10,14,17] (also reported as 3.162 Å – 3.23 Å[12,13,18–20]).† $ZnGeN_2$ is particularly interesting because it is nearly lattice-matched to GaN (0.9% mismatch) and is theorized to have a cation-disorder tunable band gap, giving access to optoelectronics with lattice-matched active layers in the ~ 2.6 eV – 3.4 eV range.[21] GaN, already active in the commercial sector, then provides an established base technology for $ZnGeN_2$ devices.

---

† For structures reported as the cation-ordered orthorhombic supercell, the wurtzite lattice constant is $a_w$ = b/2 for structures referenced to $Pnb2_1$ and $a_w$ = a/2 for structures referenced to $Pna2_1$, where the structure reference depends on the coordinate system chosen. Structures reported as monoclinic (close to hexagonal with γ slightly less than 120°) can be indexed to the orthorhombic structure for comparison.[16,17]

In thin-film form, ZnGeN$_2$ has been previously deposited on several substrates by metalorganic chemical vapor deposition (MOCVD),[18,22–24] hydride vapor-phase epitaxy (HVPE),[19] and sputtering.[12,21] ZnSn$_x$Ge$_{1-x}$N$_2$ alloys have also been grown by molecular beam epitaxy (MBE) with Sn content as low as x = 0.1.[25] MBE has previously been used to produce GaN and InGaN devices of the highest quality,[26,27] and demonstrating high-quality ZnGeN$_2$ by MBE will enable research into a new set of hybrid III-N/II-IV-N$_2$ devices. This work represents the first published MBE growth of ZnGeN$_2$.

Experimental

ZnGeN$_2$ thin films were epitaxially grown using a Riber compact 21 MBE system from standard Knudsen Zn and Ge effusion sources and an RF plasma nitrogen source operating at 13.56 MHz. Films were grown on 25 nm PVD AlN templates which are grown on c-plane sapphire and supplied by Kyma Technologies. Substrates are backside coated with 1.5 μm of sputtered Ta to facilitate the absorption of IR radiation from the substrate heater, and during this step the epitaxial surface is protected with photoresist. The substrates were then cleaned using a solvent rinse of Acetone/Methanol/2-Propanol followed by two sequential 10 minute etches in 4:1 H$_2$SO$_4$:H$_2$O$_2$ at 140 °C, followed by a deionized water rinse. Although GaN is the most appropriate substrate choice for ZnGeN$_2$ growth due to the near lattice match, AlN was chosen for initial studies to prevent conflicting structural and optical analysis due to the similarities between ZnGeN$_2$ and GaN. The surface of the growing sample was monitored by reflection high energy electron diffraction (RHEED) using a 20 keV Staib Instruments electron gun and a CCD camera monitoring the phosphor screen (see the SI for a representative RHEED image of the AlN template surface). Images were analyzed using KSA400 software, and the relationship between pixel spacing and reciprocal lattice space was calibrated at room temperature using a clean (111) Si standard with a visible 7 x 7 surface reconstruction.

For ZnGeN$_2$ films nucleated on GaN, GaN buffers were grown by the metal-modulated epitaxy (MME) method on the above-described AlN templates.[28,29] In this method GaN is grown at a relatively low substrate temperature (600 °C) in metal rich conditions (III/N = 1.8) under constant nitrogen plasma flux while periodically modulating the Ga shutter to allow consumption of excess Ga, reducing the total dislocation density and providing smooth surfaces as evidenced by RHEED surface reconstructions. For all samples grown on GaN buffers, the GaN layer was grown to at least 400 nm and until a surface reconstruction was apparent rather than to a specific thickness. A representative RHEED image of the reconstructed GaN buffer surface is shown in the supplementary info. ZnGeN$_2$ films were then grown on these GaN buffers without breaking vacuum. For ZnGeN$_2$ layers, the Zn flux was set to either $3.6 \times 10^{-6}$ torr beam equivalent pressure (BEP) or $5 \times 10^{-6}$ torr BEP for a Zn:Ge ratio of either 100 or 170. All films were grown extremely Ge limited to promote Zn incorporation at high substrate temperatures with Ge:N ratios of 0.04 – 0.1 and overall Ge-fluxes of $3.2 – 3.9 \times 10^{-8}$ torr BEP. The N flux was determined in units of stoichiometric Ga flux during GaN growth by measuring the Ga-metal consumption time while monitoring the RHEED transient.[29]

X-ray diffraction (XRD) analysis was performed using a Panalytical MRD Pro emitting Cu Kα radiation and equipped with an incident beam hybrid monochromator – a Göbel mirror plus a 4-bounce Ge (400) monochromator. The diffracted beam was analyzed with a 3-bounce Ge (220) monochromator. Cation composition was interrogated using a benchtop Fischer XDV-SDD X-ray fluorescence (XRF) system, assuming that all cations are either Zn or Ge, while using a 2-layer model to account for the GaN buffer layer. Scanning electron microscopy (SEM) images were captured using a Hitachi S4800 field emission system at 3 kV, 10 μA, and a working distance of 8.4 mm – 8.7 mm. Tapping mode atomic force microscopy (AFM) was performed on an Asylum Research system using a tip with a 150 N·m force constant to characterize surface morphology.

A cross-section transmission electron microscopy (TEM) specimen was prepared using the FIB liftout technique with the final Ga+ ion milling performed at 3 kV. Ga+ ion FIB damage was subsequently removed using low energy (< 1kV) Ar+ ion milling in a Fischione Nanomill with the sample cooled by liquid nitrogen. TEM imaging and selected area electron diffraction (SAED) analysis were performed in a FEI Tecnai SuperTwin TEM operated at 300-kV. High resolution TEM imaging and scanning transmission electron microscopy (STEM) imaging and energy dispersive x-ray spectrometry (EDS) analysis were performed in a FEI Tecnai F20 UltraTwin field emitting gun STEM operated at 200 kV.

Results & Discussion

$ZnGeN_2$ thin films were initially grown directly on AlN templates to allow unambiguous characterization of resulting films. GaN is an ideal substrate for $ZnGeN_2$, but these two materials are structurally and optically similar enough ($\Delta c \approx 0.3$ Å and $\Delta E_g < 0.2$ eV) that deconvolving the effects of GaN and $ZnGeN_2$ is difficult. However, it was discovered that growth of $ZnGeN_2$ on AlN templates only occurs within a narrow growth window in substrate temperature. Figure 1 shows the RHEED patterns of films grown at substrate temperatures ranging from 215 °C - 350 °C. Growth above 225 °C was necessary to crystallize the material but resulted in sub-stoichiometric Zn content in the films as measured by XRF, coincidentally roughening the film and introducing weak crystalline and polycrystalline features to the RHEED patterns. The film grown at 215 °C is stoichiometric in Zn content (see Figure 2) but is polycrystalline due to insufficient adatom diffusion. The films grown at higher temperatures are more crystalline due to increased adatom diffusion, but increased Zn desorption leads to sub-stoichiometric cation compositions and poor crystallinity evidenced by weak and diffuse RHEED patterns.

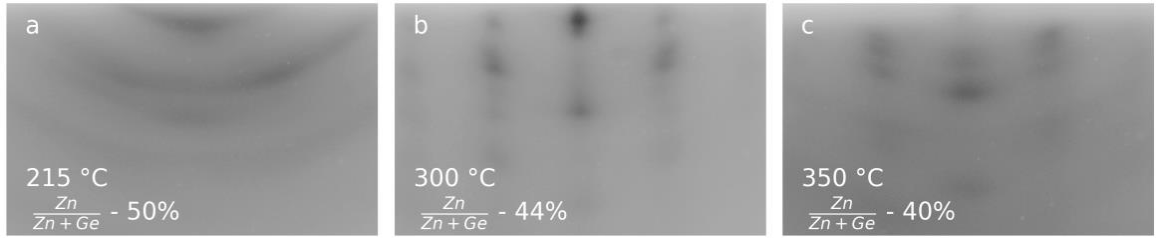

Figure 1. RHEED patterns of ZnGeN$_2$ films grown on AlN templates at 215 °C (a), 300 °C (b) and 350 °C (c). Large diffuse spots indicate poor crystallinity and rough surfaces, and the broadening of spots in to arcs represents grain misorientation trending towards polycrystalline nature.

It was therefore determined that a narrow growth window exists for epitaxial films limited by adatom diffusion at low substrate temperatures and Zn desorption at high temperatures. The relationship between Zn content and substrate temperature is shown in Figure 2, with films grown on AlN templates shown in blue. Comparing this result to the net surface Zn flux (shown as a grey curve in Figure 2), calculated as the supplied Zn flux (5 × 10$^{-6}$ torr BEP) divided by the temperature dependent vapor pressure[30], it is observed that sub-stoichiometric Zn incorporation on AlN templates occurs at approximately the same temperature that the net surface Zn flux drops below unity (Figure 2 top axes). This indicates that the net positive Zn desorption rate is faster than the crystallization rate at these temperatures, leading to poorly crystallized and Zn-deficient films. Based on the weak and diffuse RHEED patterns observed for films sub-stoichiometric in Zn, it is hypothesized that the films are phase separated into highly textured but polycrystalline ZnGeN$_2$ in an amorphous matrix of Ge$_3$N$_4$. This hypothesis is supported by TEM images of films grown at high temperatures containing less than 20% Zn cations, measured as Zn/(Zn+Ge), which show completely amorphous films primarily composed of Ge and N (see Supporting Information).

GaN buffers were investigated as a method to reduce the necessary adatom diffusion lengths and energy barriers to crystallization due to high quality surfaces and close lattice match (< 1% mismatch). To compare the effects of GaN and AlN surfaces on the nucleation of ZnGeN$_2$, the GaN buffer was grown for at least 400 nm, sometimes thicker, and until a satisfactory surface reconstruction was visible. GaN buffers were nucleated directly on AlN without use of a low temperature nucleation layer to more directly compare films of similar total dislocation density—

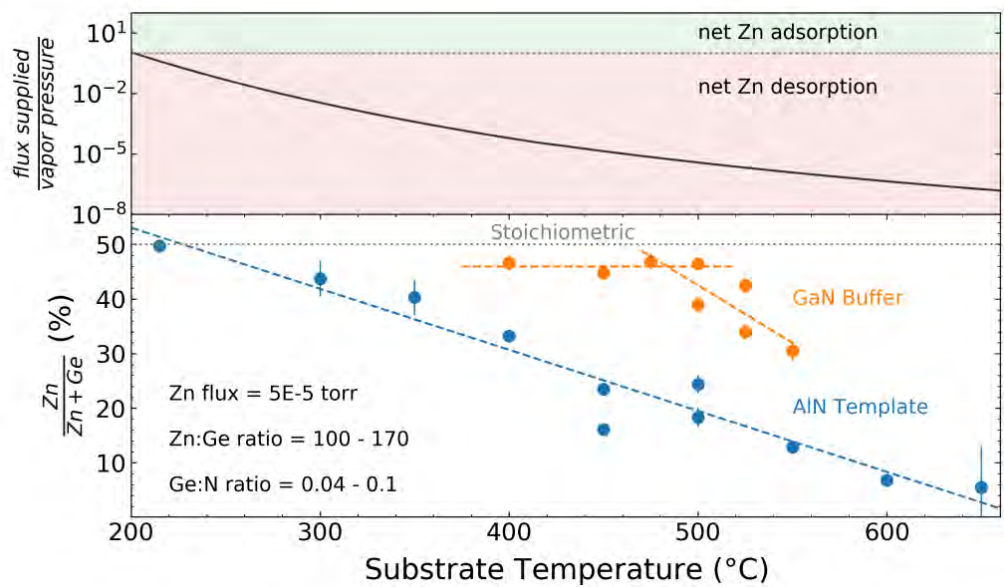

Figure 2. (Bottom axes) Zn cation composition as a percent of total (Zn + Ge) cations versus substrate temperature as measured by XRF. Films grown on AlN templates without a buffer layer incorporate less Zn at a given substrate temperature than films grown on a GaN buffer. The Zn incorporation in the film drops below the stoichiometric level at approximately the same temperature at which the net Zn flux (top axes) drops below unity for films without a buffer layer. Films grown on a buffer layer do not incorporate excess Zn, but rather saturate at a slightly Zn-poor condition of ~46%. Error bars indicate measured variation across the sample. Vapor pressure curve re-created from CRC tables.[30]

GaN and ZnGeN$_2$ have similar lattice constants, therefore the use of a low dislocation GaN layer should result in a lower dislocation density ZnGeN$_2$ film, making the surface energy comparison more difficult. The GaN buffer introduces an increased surface roughness, 1.4 nm RMS compared to 0.4 nm RMS for the AlN template (see SI Figure 2), however the use of this buffer is shown to significantly improve the quality of the ZnGeN$_2$ epilayers.

Films grown on GaN buffers are compared to films grown directly on AlN templates in Figure 2. For films grown at identical substrate temperatures, those grown on GaN buffers have significantly higher Zn content. In contrast to amorphous films grown on AlN templates at low temperatures (< 200 °C), the films grown on GaN buffers do not incorporate super-stoichiometric concentrations of Zn; instead, the Zn-content of the film saturates around 46 % – 47 %. This is a reasonable observation, as both the mobility and vapor pressure of Zn are relatively high at the growth temperatures used, which could lead to sub-stoichiometric Zn even at ideal growth conditions. It is also within the detection resolution of the XRF tool used which has a noise floor of approximately 3%, meaning that higher resolution methods will be necessary to interrogate stoichiometry on a finer level.

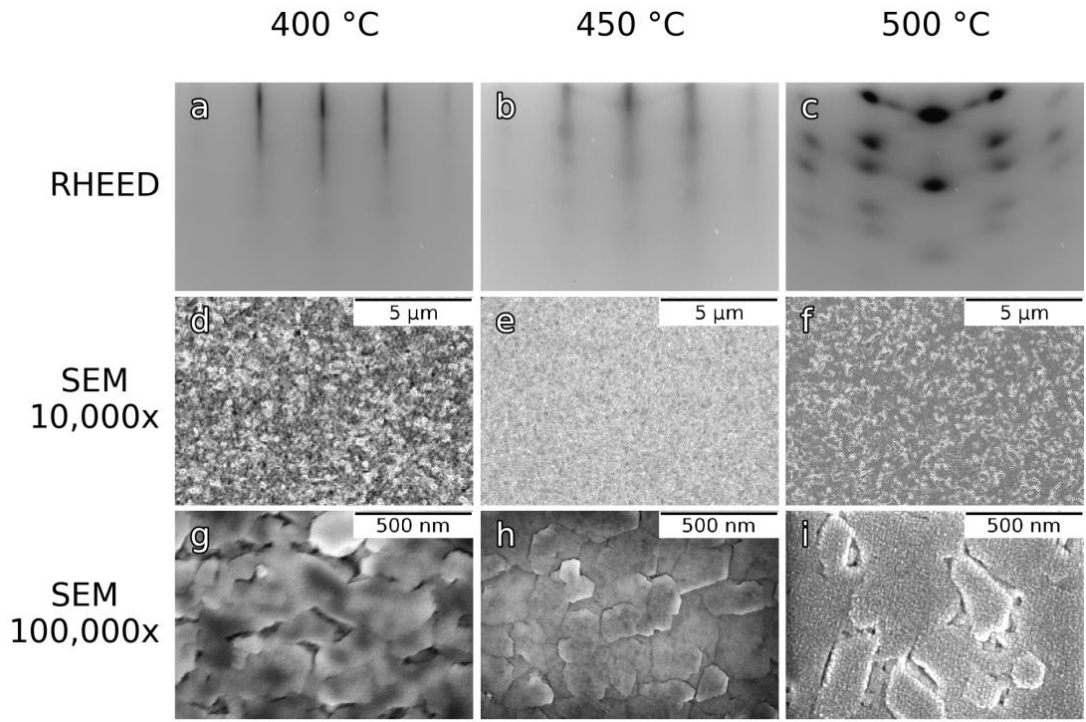

Figure 3. RHEED patterns (a – c), and 10,000x (d-f) and 100,000x (g-i) SEM images of ZnGeN$_2$ thin films grown on GaN buffers at 400 °C (a,d,g), 450 °C (b,e,h), and 500 °C (c,f,i). Films grown at 500 °C exhibit a broad, diffuse, spotty RHEED pattern exhibiting high roughness with streaks characteristic of faceting. Corresponding SEM images show the surface microstructures responsible for the faceting signal in RHEED. Films grown at 450 °C have smaller grains and with faceting reduced but not eliminated. Films grown at 400 °C show no faceting in RHEED or SEM.

For the ZnGeN$_2$ films on GaN buffers grown close to stoichiometry, it is observed that decreasing substrate temperature significantly improves the surface morphology and crystallinity of the growing film, as shown in Figure 3. From left to right, the columns represent films grown at 400 °C, 450 °C, and 500 °C. The 500 °C film shows a very rough RHEED pattern (Figure 3c) with strong faceting and broad spots. The structure of the RHEED pattern is similar to those grown on AlN; however, the significantly increased spot intensity indicates an increase in film crystallinity. The microstructure responsible for the faceting is visible by SEM at 100,000x magnification (Figure 3i) as crystallites protruding from the surface of the larger grains. Similar features are observed on films nucleated directly on AlN at similar growth temperatures (see SI figure 3). Decreasing the substrate temperature to 450 °C significantly improves the crystallinity and surface morphology as observed by RHEED (Figure 3b), with the onset of a streaky pattern superimposed on top of broad spots and faceting which is diminished but not eliminated. The SEM images above (Figure 3e and h) corroborate the RHEED image showing a near elimination of the observed surface faceting and improved coalescence, with only a few nanocrystals visible above the smooth grains. The grain size of this 450 °C film is also reduced, as is expected for lower temperature growth. Finally, the film grown at 400 °C shows an elimination of faceting in both RHEED (Figure 3a) and SEM (Figure 3d and g).

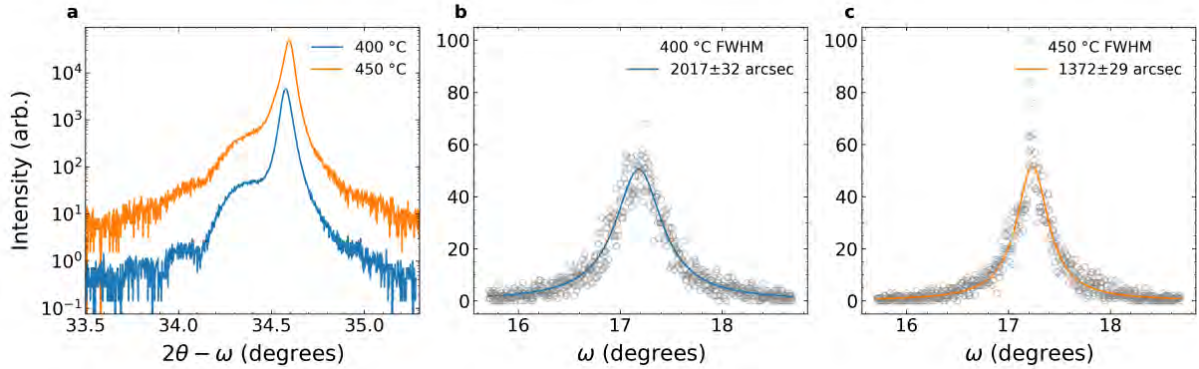

Figure 4. (a) X-ray diffraction coupled scans of ZnGeN$_2$ film grown at 400 °C and 450 °C on MME GaN buffers. (b-c) Omega scans taken at 2θ = 34.4° showing a significant increase in crystal quality by increasing the substrate temperature from 400 °C (b) to 450 °C (c). Scans were acquired using a triple-crystal monochromator on the diffracted beam path to exclude signal from the GaN layer (Δ2θ ≈ 0.15°).

High-resolution XRD was also employed to investigate the structural properties of the films. Figure 4a shows symmetric coupled 2θ-ω scans of (002) GaN and (002) disordered ZnGeN$_2$, and omega scans taken at 2θ ≈ 34.5° are shown in Figure 4b and c. The term 'rocking curve' is avoided here because the scans were acquired with a triple-crystal monochromator on the diffracted beam side to remove signal contributions from the GaN layer due to the small 2θ separation (Δ2θ = 0.14°). When using traditional diffracted beam optics, the ZnGeN$_2$ layer is only visible in symmetric coupled scans as a small shoulder, and rocking curves are dominated by the GaN layer. The omega scans indicate that the film grown at 450 °C has a significantly lower degree of c-plane tilt when compared to the film grown at 400 °C (1370 arcsec compared with 2020 arcsec). Despite this result, the film grown at 400 °C shows stronger Pendellösung thickness fringes, a measure of uniformity in the vertical coherence length (a combination of thickness uniformity, interface roughness, surface roughness, and permittivity). As discussed in the following paragraph, the film at 450 °C has an overall lower surface roughness indicating that either the film grown at 400 °C has a smoother interface compared to the film grown at 450 °C, or the dielectric constant of the 400 °C ZnGeN$_2$ layer is more uniform. The film grown at 500 °C did not show strong enough signal from the ZnGeN$_2$ layer to warrant analysis and is shown in the Supporting Information. Finally, the c- lattice parameter for the films shown in Figure 4 is determined to be 5.215 Å ± 0.005 Å.

To quantify roughness, the films shown by SEM in Figure 3 were also investigated by AFM. Figure 5 shows 10 μm x 10 μm (a – c) and 1 μm x 1 μm (d – f) scans (corresponding roughly to the 10,000x and 100,000x SEM images) of films grown on GaN buffers at 400 °C (Figure 5a and d), 450 °C (Figure 5b and e), and 500 °C (Figure 5c and f). The corresponding roughness values, in order of decreasing substrate temperature (increasing crystal quality), are measured at 5.1 nm RMS, 4.2 nm RMS, and 6.6 nm RMS. Interestingly, the ZnGeN$_2$ film grown at 450 °C exhibits the lowest roughness surface, indicating that the interplay between adatom diffusion, surface energy, and Zn desorption are also important for roughness considerations. It is also important to note that while the overall roughness of the intermediate temperature film is lower, the lowest temperature film shows the smoothest individual grains, and further studies on thicker films will most likely benefit from the use of a low temperature nucleation layer, similar to growth schemes used for the nucleation of high quality GaN on dissimilar substrates.[31] Also of note, the nanostructure responsible for the faceting on the 450 °C and 500 °C samples is visible in Figure 5e and f as ~ 20 nm – 40 nm wide features extending 1 nm – 3 nm above the surface as determined by AFM line scans.

Due to the overlapping signature of the GaN buffer and ZnGeN$_2$ thin film when measured *ex-situ* by XRD, information about the evolution of the growing ZnGeN$_2$ crystal must be obtained by RHEED. Accurate measurements of the in-plane lattice constant can be obtained by measuring the spacing of the RHEED streaks/spots, where the a-spacing of the wurtzite lattice is probed by aligning the RHEED beam parallel to $\{11\bar{2}0\}$ planes. Extracted values of the a- lattice parameter for GaN (after buffer growth at 400 °C) and ZnGeN$_2$ are reported as a function of the growing film thickness in Figure 6. Within the error given, the ZnGeN$_2$ thin film appears to grow pseudomorphically for the first 20 nm of growth, relaxing to a measured a-spacing of 3.216 Å ± 0.004 Å after 26 nm. The error shown represents the standard deviation in the measurement from a single RHEED image (measuring the RHEED streak separation at every pixel along the streak above an intensity threshold), however additional sources of error, including substrate wobble and beam focus, reduce the overall measurement accuracy to approximately ± 0.01 Å. Regardless, the critical thickness for relaxation of coherent strain for ZnGeN$_2$ on GaN (at the given growth conditions) is in the range 20 nm - 26 nm. At these thicknesses ZnGeN$_2$ quantum wells and barriers can be grown pseudomorphically within a GaN structure – an important feature for future ZnGeN$_2$-GaN hybrid LEDs.

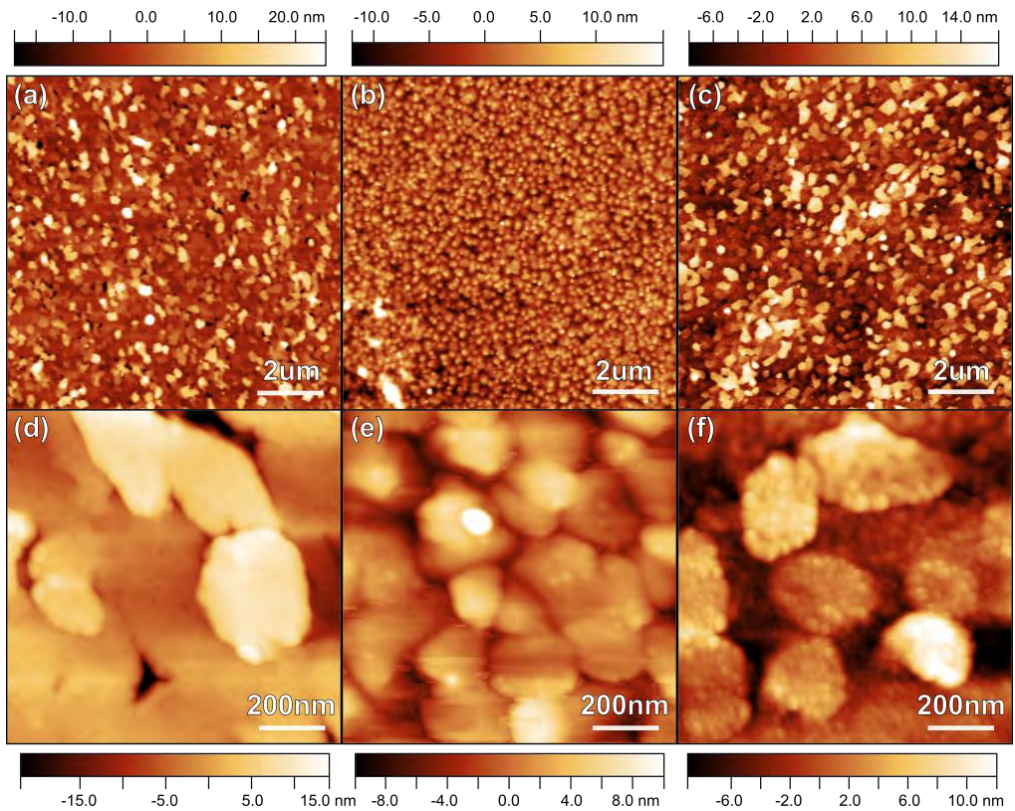

Figure 5. Atomic force micrographs of ZnGeN$_2$ thin films grown on GaN buffers at 400 °C (a and d), 450 °C (b and e), and 500 °C (c and f), corresponding to the SEM and RHEED images in Figure 3. The surface microstructure responsible for the faceting observed in RHEED is seen as ~30 nm wide crystallites on top of the ~300 nm grains.

Finally, to investigate the structure and crystallinity of the ZnGeN$_2$ layer, cross-sectional TEM and SAED were performed on a thin specimen prepared from the ZnGeN$_2$ film grown on GaN at 400 °C. Low magnification bright field TEM imaging (Figure 7a) revealed a ~40 nm thick ZnGeN$_2$ layer had grown on the GaN buffer layer. High resolution TEM imaging (Figure 7b) and SAED (Figure 7c) revealed the layer to be crystalline and epitaxially oriented on the GaN buffer layer. Threading dislocations and other defects were observed to propagate through both the GaN buffer layer and the epitaxial ZnGeN$_2$ layer grown on top. The SAED pattern (Figure 7c) obtained from an area including the ZnGeN$_2$ layer and GaN buffer layer is consistent with the SAED pattern calculated for the cation-disordered wurtzite ZnGeN$_2$ crystal structure combined with GaN (Figure 7d). Superlattice diffraction spots would be expected if the ZnGeN$_2$ formed in the cation-ordered orthorhombic phase (see the calculated diffraction patterns included in the supporting information). STEM EDS elemental maps (included in the supporting information) confirmed the growth of a uniform ZnGeN$_2$ layer on top of the GaN buffer layer.

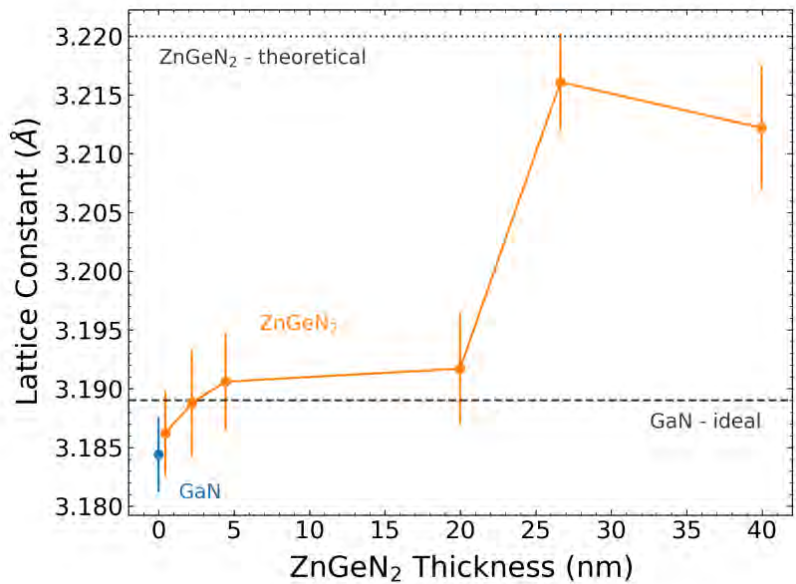

Figure 6. ZnGeN$_2$ (orange) and GaN (blue) lattice spacing as measured by RHEED. The critical thickness for ZnGeN$_2$ pseudomorphic growth is observed between 20 nm and 26 nm, where the film thickness is obtained by cross-sectional SEM.

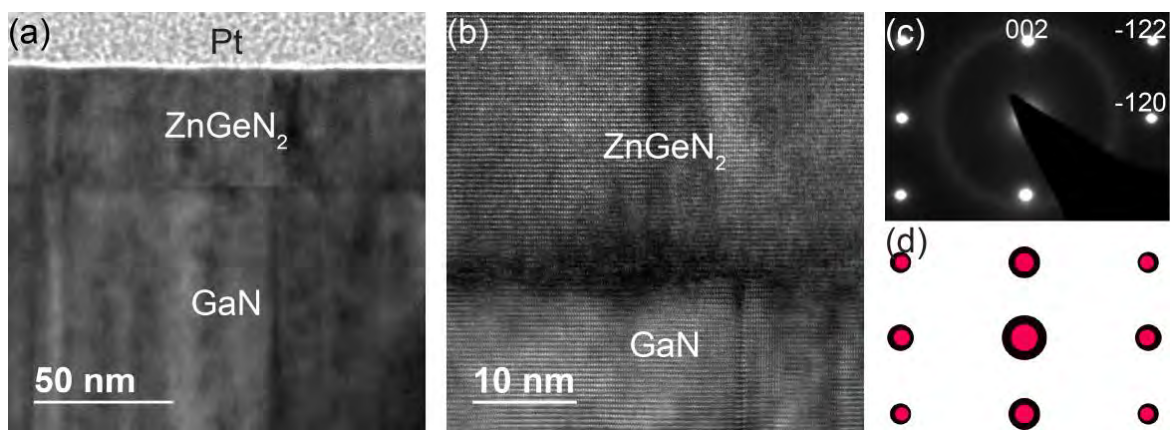

Figure 7. (a) Bright field TEM image of epitaxial ZnGeN$_2$ layer grown at 400 °C on a GaN buffer; (b) High resolution TEM image of interface; (c) SAED pattern obtained from an area including both ZnGeN$_2$ and GaN layers; (d) Calculated SAED pattern for cation-disordered ZnGeN$_2$ (red spots) overlaid on GaN (black spots) along same direction as (c).

Conclusions

ZnGeN$_2$ thin films were grown directly on AlN templates and on GaN buffers by MBE to investigate the effect of surface energy on nucleation. The nearly lattice-matched GaN surface allows greater incorporation of Zn at higher substrate temperatures, leading to significantly increased crystal quality. This effect is likely due to the lattice-matched buffer providing a lower energetic barrier to crystallization. Lowering substrate temperatures from 500 °C to 400 °C eliminates surface faceting as measured by RHEED, but AFM measured surface roughness increases below 450 °C, indicating the need for further growth window optimization. The critical thickness of ZnGeN$_2$ on GaN at the growth conditions described is determined as 20 nm – 25 nm. The relaxed lattice parameters for this film were determined by XRD and calibrated RHEED as a = 3.216 Å ± 0.004 Å and c = 5.215 Å ± 0.005 Å, comparable to literature values of a = 3.22 Å and c = 5.19 Å. This work demonstrates a significant step towards the development of hybrid ZnGeN$_2$-GaN integrated optoelectronic devices.

ASSOCIATED CONTENT

**Supporting Information.**

Figure S1: Representative RHEED patterns of AlN and GaN nucleating surfaces

Figure S2: AFM images of a representative AlN template and GaN buffer

Figure S3: TEM of amorphous Zn-poor films nucleated on AlN

Figure S4: SEM of films nucleated on AlN

Figure S5: STEM EDS maps of ZnGeN$_2$ on GaN, calculated TED pattern for ordered ZnGeN$_2$

Figure S6: Additional XRD characterization

**Notes**

The authors declare no competing financial interest.

ACKNOWLEDGMENT

The authors would like to acknowledge Bobby To for obtaining SEM images of the films described. This work was authored in part by Alliance for Sustainable Energy, LLC, the manager and operator of the National Renewable Energy Laboratory for the U.S. Department of Energy (DOE) under Contract No. DE-AC36-08GO28308. This work was supported by the U.S. Department of Energy, Office of Science, Basic Energy Sciences, Materials Sciences and Engineering Division. The views expressed in the article do not necessarily represent the views of the DOE or the U.S. Government. The U.S. Government retains and the publisher, by accepting the article for publication, acknowledges that the U.S. Government retains a nonexclusive, paid-up, irrevocable, worldwide license to publish or reproduce the published form of this work, or allow others to do so, for U.S. Government purposes.